\documentclass[journal]{IEEEtran}
\usepackage{multirow}
\usepackage{makecell}
\usepackage{amssymb}
\usepackage{graphicx}
\usepackage{algorithm}
\usepackage{algpseudocode}

\ifCLASSINFOpdf
\else
\fi
\begin{document}
%
\title{Towards Domain-Specific Cross-Corpus Speech Emotion Recognition Approach}
%
%
%

\author{Yan~Zhao,
        Yuan~Zong$^*$,~\IEEEmembership{Member,~IEEE,}
	Hailun~Lian,
	 Cheng~Lu,~Jingang~Shi,    
and~Wenming~Zheng$^*$,~\IEEEmembership{Senior~Member,~IEEE}
\thanks{Y. Zhao and H. Lian are with the Key Laboratory of Child Development and Learning Science of Ministry of Education, Southeast University, Nanjing 211189, China, and also with the School of Information Science and Engineering, Southeast University, Nanjing 211189, China.}
\thanks{Y. Zong, W. Zheng, and C. Lu are with the Key Laboratory of Child Development and Learning Science of Ministry of Education, Southeast University, Nanjing 211189, China, and also with the School of Biological Science and Medical Engineering, Southeast University, Nanjing 211189, China.}
\thanks{J. Shi is with the School of Software, Xi'an Jiao Tong University, Xi'an 710049, China.}
\thanks{$^*$ Corresponding authors.}}

%
%

\markboth{Journal of \LaTeX\ Class Files,~Vol.~14, No.~8, August~2015}%
{Shell \MakeLowercase{\textit{et al.}}: Bare Demo of IEEEtran.cls for IEEE Journals}
%



\maketitle

\begin{abstract}
Cross-corpus speech emotion recognition (SER) poses a challenge due to feature distribution mismatch, potentially degrading the performance of established SER methods. In this paper, we tackle this challenge by proposing a novel transfer subspace learning method called acoustic knowledge-guided transfer linear regression (AKTLR). Unlike existing approaches, which often overlook domain-specific knowledge related to SER and simply treat cross-corpus SER as a generic transfer learning task, our AKITR method is built upon a well-designed acoustic knowledge-guided dual sparsity constraint mechanism. This mechanism emphasizes the potential of minimalistic acoustic parameter feature sets to alleviate classifier over-adaptation, which is empirically validated acoustic knowledge in SER, enabling superior generalization in cross-corpus SER tasks compared to using large feature sets. Through this mechanism, we extend a simple transfer linear regression model to AKTLR. This extension harnesses its full capability to seek emotion-discriminative and corpus-invariant features from established acoustic parameter feature sets used for describing speech signals across two scales: contributive acoustic parameter groups and constituent elements within each contributive group. Our proposed method is evaluated through extensive cross-corpus SER experiments on three widely-used speech emotion corpora: EmoDB, eNTERFACE, and CASIA. The results confirm the effectiveness and superior performance of our method, outperforming recent state-of-the-art transfer subspace learning and deep transfer learning-based cross-corpus SER methods. Furthermore, our work provides experimental evidence supporting the feasibility and superiority of incorporating domain-specific knowledge into the transfer learning model to address cross-corpus SER tasks.
\end{abstract}


%
\IEEEpeerreviewmaketitle

\section{Introduction}

\IEEEPARstart{S}{peech} plays a crucial role in human daily communication, serving as a natural means for individuals to express their emotions such as \textit{Happiness}, \textit{Fear}, and \textit{Sadness}. As a result, the research of speech emotion recognition (SER)~\cite{akccay2020speech,singh2022systematic,wagner2023dawn}, which seeks to empower computers to automatically understand emotional states from speech signals, holds significant practical value. Over the past few decades, SER has garnered substantial attention within the communities of human-computer interaction, affective computing, and signal processing, leading to the development of numerous well-performing SER methods~\cite{nwe2003speech,huang2014speech,fayek2017evaluating,lu2022speech,lu2022domain,zhang2022spontaneous}.  

However, it is important to note that most established SER methods, including those mentioned above, primarily focus on an ideal scenario where the training and testing speech signals belong to the same speech emotion corpus. In practical situations, the testing speech signals may differ significantly from the training speech signals, exhibiting variations in numerous factors, such as languages, recording equipment, and environmental conditions. This gives rise to a challenging but intriguing task known as \textit{cross-corpus SER}~\cite{schuller2010cross} within the field of SER. In cross-corpus SER tasks, the training and testing speech signals originate from different speech emotion corpora and can be referred to as the source and target signals, respectively. Moreover, while we have access to ground truth emotion labels for the source speech samples, the target speech emotion corpus remains entirely unlabeled. 

In the early stages, the research of cross-corpus SER mostly focus on feature engineering, aiming to enhance the corpus-invariant ability of acoustic parameter feature sets used to describe speech signals. For example, in the work of~\cite{schuller2010cross}, three feature normalization schemes, including corpus normalization, speaker normalization, and speaker-corpus normalization, are designed to address feature distribution mismatches between source and target speech emotion corpora. Subsequently, Parlak et al.~\cite{parlak2014cross} attempt to use numerous feature selectors, such as linear forward selection, to seek high-quality speech features that are robust to corpus variance from existing comprehensive acoustic feature sets. In recent years, inspired by the tremendous success of transfer learning in various cross-domain recognition tasks~\cite{pan2009survey,niu2020decade}, researchers have shifted their focus to the development of transfer learning methods for cross-corpus SER. These methods have achieved promising performance in recognizing emotions in speech signals across different corpora, marking a significant advancement in this field. 

Broadly, current transfer learning-based cross-corpus SER methods can be classfied into two types, including \textit{Transfer Subspace Learning} and \textit{Deep Transfer Learning}:

(1) Transfer subspace learning-based cross-corpus SER methods typically begin by using a set of acoustic low-level descriptors (LLDs), such as fundamental frequency (F0) and Mel-frequency cepstral coefficients (MFCC), along with their associated functions, such as maximal and mean values, to describe the source and target speech signals. Subsequently, a transfer subspace learning model is developed to mitigate the distribution mismatch between the two feature sets. One early method can be traced back to the work of~\cite{hassan2013acoustic}, in which Hassan et al. extend the support vector machine (SVM)~\cite{cortes1995support} to an importance-weighted SVM (IW-SVM) for cross-corpus SER. IW-SVM incorporates three different transfer subspace learning models: kernel mean matching (KMM)~\cite{gretton2009covariate}, unconstrained least-squares importance fitting (uLSIF)~\cite{kanamori2009least}, and Kullback-Leibler importance estimation procedure (KLIEP)~\cite{tsuboi2009direct}. It is hence enabled to learn a set of weights for the source speech samples, ensuring that the weighted source speech feature sets align with the distribution of target speech feature sets. Another notable work is the transfer non-negative matrix factorization (TNNMF) models designed by Song et al.~\cite{song2016cross}. These models integrate the maximum mean discrepancy (MMD)~\cite{borgwardt2006integrating} to measure and minimize the discrepancies between the source and target speech feature distributions. Following this work, Luo et al.~\cite{luo2020nonnegative} further advance TNNMF model by jointly reducing the marginal and class-aware conditional feature distribution gaps between the two different speech sample sets.

(2) In contrast to transfer subspace learning, deep transfer learning methods often utilize the speech spectrums of the original speech signals as input for deep neural networks, harnessing their powerful nonlinear representation capabilities to learn emotion-discriminative and corpus-invariant features. Parry et al.~\cite{parry2019analysis} examine the generalization capacity of deep neural networks for cross-corpus SER across six different speech emotion corpora. Their experimental results demonstrate that convolutional neural networks (CNNs)~\cite{krizhevsky2012imagenet} exhibit superior generalisation capabilities compared to recurrent neural networks (RNNs)~\cite{hochreiter1997long}. Insipired by this observation, Zhao et al.~\cite{zhao2022deep} propose deep transductive transfer regression networks (DTTRN) based on CNN architectures. A key contribution of DTTRN is the incorporation of additional fine-grained emotion class-aware conditional MMD, which aids in better bridging the distribution gap between learned source and target features compared to the original MMD. Additionally, Zhao et al.~\cite{zhao2023deep} introduce another CNN-based deep transfer learning method called deep implicit distribution alignment neural networks (DIDAN). DIDAN performs implicit distribution alignment for source and target speech corpora by replacing the minimization of MMD with sparsely reconstructing target samples using source samples. More recently, domain-adversarial learning-based models~\cite{gideon2019improving,gao2022domain,gao2023adversarial} have been developed to learn more generalized representations of speech signals for cross-corpus SER. The key concept behind these methods are the introduction of an additional domain (corpus) classifier, which enables the deep neural networks to learn the generalized features to describe speech signals, regardless of their corpus sources.

While both transfer subspace learning and deep transfer learning methods have demonstrated success in addressing the challenge of cross-corpus SER, it is worth noting that these methods often approach cross-corpus SER as a generic transfer learning task. This means that most of these methods focus primarily on developing transfer learning models without specifically considering the valuable acoustic knowledge inherent to SER. As a result, these transfer learning methods can be applied to other cross-domain recognition tasks without making significant modifications. According to the "No Free Lunch Theorem"~\cite{wolpert1997no}, it is established that \textit{"There is no universal learning algorithm that can provide the best solution for every problem. Each algorithm has its strengths and weaknesses, and its performance is highly dependent on the specific problem domain and data distribution."} From this perspective, it can be argued that they may not offer ultimately satisfactory solutions for cross-corpus SER. In other words, incorporating domain-specific knowledge from SER to guide the design of transfer learning models could potentially lead to even better performance compared to the generic transfer learning models when dealing with cross-corpus SER. Therefore, our goal in this paper is to develop a domain-specific transfer learning approach for cross-corpus SER. Specifically, we propose a novel transfer subspace learning method called acoustic knowledge-guided transfer linear regression (AKTLR).

\begin{figure}[t!]
\centering
\includegraphics[width=0.95\columnwidth]{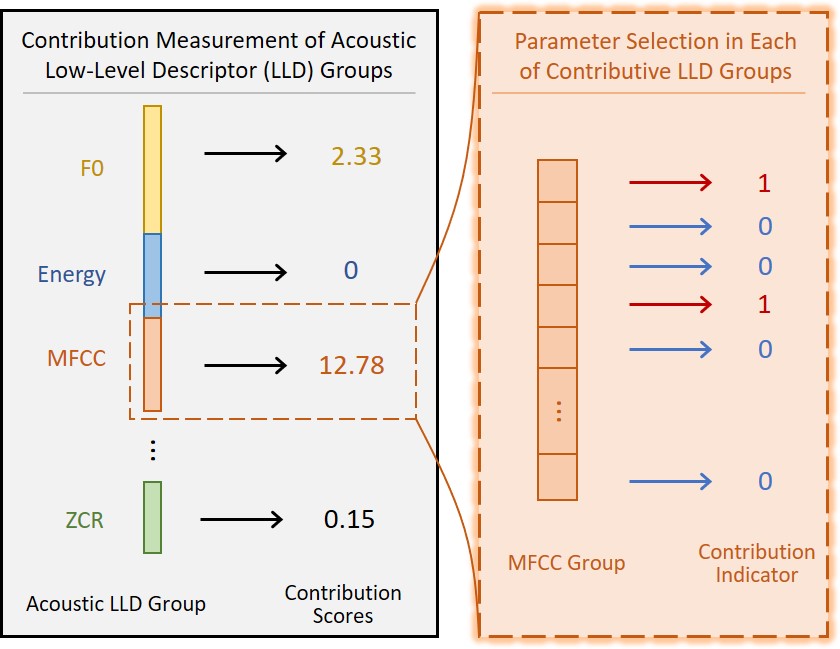}
\caption{Acoustic Knowledge-Guided Dual Sparsity Constraint Mechanism: The Concept behind the Proposed AKTLR Method for Addressing Cross-Corpus SER Tasks.}
\label{fig:nb1}
\end{figure}

The basic concept behind AKTLR comes from the empirically validated acoustic knowledge about the acoustic parameter feature sets designed for describing speech signals and their cross-corpus recognition performance evaluation within SER~\cite{williams1972emotions,parlak2014cross,eyben2015geneva}. These works inform us that selectively minimalistic high-quality acoustic parameters are more capable of exhibiting superior generalization ability to variance in speech emotion corpus.Therefore, selecting these acoustic parameters may enable the transfer subspace learning models to achieve more promising recognition performance in cross-corpus SER tasks compared to directly using larger feature sets comprising comprehensive acoustic parameters. This insight motivates us to introduce an acoustic knowledge-guided dual sparsity constraint mechanism, illustrated in Fig.~\ref{fig:nb1}, to develop AKTLR model for cross-corpus SER. As depicted in Fig.~\ref{fig:nb1}, this mechanism equips the AKTLR model to proficiently discern emotion-discriminative and corpus-invariant features from established acoustic parameter feature sets at both coarse-grained and fine-grained scales. Specifically, it begins by measuring the contribution scores of different acoustic LLDs, and subsequently selects truly contributive derived features from each of LLD groups with high contribution scores.


To evaluate the effectiveness of AKTLR, we conduct extensive cross-corpus SER experiments using three widely-used speech emotion corpora: EmoDB~\cite{burkhardt2005database}, eNTERFACE~\cite{martin2006enterface}, and CASIA~\cite{zhang2008design}. The experimental results demonstrate that our AKTLR outperforms recent state-of-the-art transfer subspace learning and deep transfer learning-based cross-corpus SER methods, showcasing the effectiveness of incorporating domain-specific knowledge into transfer subspace learning for cross-corpus SER. In summary, this paper makes three primary contributions:
\begin{enumerate}
\item We propose AKTLR, a novel transfer subspace learning method inspired by empirically verified acoustic knowledge, making it the first work to propose a domain-specific approach for cross-corpus SER.
\item We introduce an acoustic knowledge-guided dual sparsity constraint mechanism to guide the design of AKTLR. This mechanism enables AKTLR to effectively seek emotion-discriminative and corpus-invariant features from established acoustic parameter feature sets, operating at two different scales, for cross-corpus SER.
\item We perform extensive cross-corpus SER experiments using three widely-used speech emotion corpora to assess the effectiveness of AKTLR. The experimental results demonstrate the superior performance of AKTLR in addressing the challenge of cross-corpus SER.
\end{enumerate}

The subsequent sections of this paper are structured as follows: Section~\ref{sec:proposedmethod} provides detailed explanations of the proposed AKTLR method. In Section~\ref{sec:experiments}, we evaluate the performance of the AKTLR method in tackling the challenge of cross-corpus SER. Finally, the paper is concluded in Section~\ref{sec:conclusion}.

\section{Proposed Method}
\label{sec:proposedmethod}
\subsection{Notations}

In this section, we will provide a detailed description of the proposed AKTLR model and demonsrtate how to utilize this model to address cross-corpus SER tasks. Before delving into the model specifics, let us establish a set of notations necessary for constructing the model. Suppose we have a source speech emotion corpus comprising $N_s$ samples, with its feature matrix denoted as $\mathbf{X}^s = [{\mathbf{X}_1^s}^T,\cdots,{\mathbf{X}_{G}^s}^T]^T\in \mathbb{R}^{d\times N_s}~(d = \sum_{i=1}^{G} d_i)$. Here, $N_s$ represents the dimension of acoustic parameter feature vector and $d$ represents the feature dimension. $\mathbf{X}_i^s \in \mathbb{R}^{d_i\times N_s}$ represents the features derived from a LLD group (one specific LLD or more closely-related LLDs), such as MFCC or energy-based features, within $G$ LLD groups used to design acoustic parameter features for describing speech signals. The corresponding emotion label matrix of the source speech samples is expressed as $\mathbf{Y}_s = [\mathbf{y}_1^s,\cdots,\mathbf{y}_{N_s}^s] \in \mathbb{R}^{C\times N_s}$. Each column $\mathbf{y}_j^s = [y_{j,1},\cdots,y_{j,C}]^T$ is a one-hot vector associated with the $j^{th}$ speech sample. The $k^{th}$ entry in $\mathbf{y}_j^s$ is set as 1 if the corresponding speech sample expresses the $k^{th}$ emotion within emotion set $\{1,\cdots,C\}$, and 0 otherwise. Similarly, the target speech feature matrix can be denoted as $\mathbf{X}^t = [{\mathbf{X}_1^t}^T,\cdots,{\mathbf{X}_{G}^s}^T]^T\in \mathbb{R}^{d\times N_t}$, where $N_t$ is number of samples in the target speech emotion corpus.

\subsection{AKTLR Model}

As previously described and illustrated in Fig.~\ref{fig:nb1}, our AKTLR method is designed based on a simple transfer linear regression model and an acoustic knowledge-guided dual sparsity constraint mechanism. This design enables the model to effectively seek high-quality speech features that are emotion-discriminative and corpus-invariant at two scales within a comprehensive feature set consisting of various acoustic parameters and their derived features. This facilitates the connection of emotions expressed in speech signals from different corpora. To achieve this, we design the following optimization problem for AKTLR:
\begin{eqnarray}
	\min_{\mathbf{P},\alpha} \mathcal{L}_{tlr} + \mu \mathcal{L}_{ds},~s.t.,~\alpha \succeq 0.
\label{eqn:nb1}
\end{eqnarray}

In Eq.(\ref{eqn:nb1}), $\mathcal{L}_{tlr}$ and $\mathcal{L}_{ds}$ represent the loss functions corresponding to a simple \textit{Transfer Linear Regression} model and newly designed \textit{Acoustic Knowledge-guided Dual Sparsity Constraint Mechanism}, respectively. The parameter $\mu$ serves as a trade-off parameter that controls the balance between these two functions. It is important to note that $\mathbf{P}\in \mathbb{R}^{C\times d}$ is the regression coefficient matrix to be learned in AKTLR, and $\alpha = [\alpha_1,\cdots,\alpha_G]^T$ is a contribution score vector and also a model parameter of AKTLR. Each entry in this vector, $\alpha_i$, is a non-negative value and measures the contribution of its corresponding acoustic parameter feature derived from a LLD group, in recognizing emotions across speech corpora. In what follows, we describe the details of the key loss functions in AKTLR.

\subsubsection{Loss Function for Transfer Linear Regression}

The loss function corresponding to transfer linear regression, denoted as $\mathcal{L}_{tlr}$, can be formulated as follows:
\begin{eqnarray}
	\mathcal{L}_{tlr} = \Vert \mathbf{Y}^s - \sum_{i=1}^{G}\alpha_i\mathbf{P}_i\mathbf{X}_i^{s}\Vert_F^2 + \lambda_1 \Vert\sum_{i=1}^{G}\alpha_i\mathbf{P}_i\Delta{\bar\mathbf{x}}_i^{st}\Vert^2,
\label{eqn:nb2}
\end{eqnarray}
where $\mathbf{P} = [\mathbf{P}_1^T,\cdots,\mathbf{P}_G^T]^T~(\mathbf{P}_i\in\mathbb{R}^{C\times d_i})$, $\Delta{\bar\mathbf{x}}_i^{st} = \frac{1}{N_s}\mathbf{X}^s\mathbf{1}_{N_s} - \frac{1}{N_t}\mathbf{X}^t\mathbf{1}_{N_t}$ is the mean difference between the source and target speech feature vectors associated with the $i^{th}$ LLD group, and $\lambda_1$ is the trade-off parameter. 

The loss function $\mathcal{L}_{tlr}$ consists of two main terms. The first term, $\Vert \mathbf{Y}^s - \sum_{i=1}^{G}\alpha_i\mathbf{P}_i\mathbf{X}_i^{s}\Vert_F^2$, represents a weighted linear regression function that establishes the relationship between the source speech feature sets and their ground truth emotion labels. Minimizing this term enables the proposed AKTLR to seek a subspace to distinguish different emotions expressed in speech signals. The second term, $\Vert\sum_{i=1}^{G}\alpha_i\mathbf{P}_i\Delta{\bar\mathbf{x}}_i^{st}\Vert^2$, measures the distribution gap between the source and target speech feature sets in such subspace using the one-order statistical moment, the mean value. Minimizing this term encourages the source and target feature sets to have similar feature distributions in such subspace. Thus, the proposed AKTLR is also applicable to distinguish emotions expressed in target speech signals.

\subsubsection{Loss Function for Acoustic Knowledge-guided Dual Sparsity Constraint Mechanism}

The loss function for the acoustic knowledge-guided dual sparsity constraint mechanism is designed as follows:
\begin{eqnarray}
	\mathcal{L}_{ds} = \Vert \alpha \Vert_1+ \tau \sum_{i=1}^{G}\Vert \mathbf{P}_i \Vert_{2,1}~(\alpha \succeq 0).
\label{eqn:nb3}
\end{eqnarray}

Here, $\tau$ is the trade-off parameter. This loss function consists of two major terms: the $l_1$ norm with respect to $\alpha$ and the $l_{2,1}$ norm with respect to $\mathbf{P}_i$. Minimizing this loss function enforces the proposed AKTLR to learn a non-negative sparse $\alpha$ and column-sparse $\mathbf{P}_i$. The non-negative sparse $\alpha$ allows the AKTLR model to measure the specific contributions of different acoustic parameter features at a coarse-grained scale of LLD group, suppressing the less-contributive ones, while highlighting highly-contributive ones. Additionally, the column-sparse $\mathbf{P}_i$ further enhance AKTLR by performing fine-grained feature selection to suppress low-quality acoustic parameter features derived from LLD groups with high contribution scores.

\subsubsection{Optimization Problem of AKTLR}

By incorporating the formulations of the two loss functions as shown in Eqs.(\ref{eqn:nb2}) and (\ref{eqn:nb3}) into Eq.(\ref{eqn:nb1}), we can derive the ultimate optimization problem for training the proposed AKTLR models, which is expressed as follows:
\begin{eqnarray}
	\min_{\mathbf{P}_i,\alpha}\Vert \mathbf{Y}^s - \sum_{i=1}^{G}\alpha_i\mathbf{P}_i\mathbf{X}_i^{s}\Vert_F^2 + \lambda_1 \Vert\sum_{i=1}^{G}\alpha_i\mathbf{P}_i\Delta\bar{\mathbf{x}}_i^{st}\Vert^2~~~~\nonumber\\+ \lambda_2 \Vert \alpha \Vert_1+ \lambda_3 \sum_{i=1}^{G}\Vert \mathbf{P}_i \Vert_{2,1},\nonumber\\ s.t.~\alpha \succeq 0.~~~~~~~~~~~~~~~~~~~~~~~~~~~~~~~~~~~~~~~~~~~~~~~~
\label{eqn:nb4}
\end{eqnarray}
where $\lambda_1$, $\lambda_2 = \mu$, and $\lambda_3 = \mu\times\tau$ are the trade-off parameters that control the balance among the key terms in the total loss function of AKTLR.

\begin{algorithm}[!t]
\caption{Updating Procedures for Learning the Optimal $\mathbf{P}$ in Eq.(\ref{eqn:nb8}).}
\label{alm:nb1}
(1) \textit{Fix $\mathbf{P}$, $\mathbf{T}$, $\kappa$, and Minimize $\mathcal{L}$ w.r.t. $\mathbf{Q}$}: This step is equivalent to solving the following optimization problem:
\begin{eqnarray}
	\min_{\mathbf{Q}} \Vert \mathbf{Y} - \mathbf{Q}\mathbf{X}\Vert_F^2 - Tr(\mathbf{T}^T\mathbf{Q}) + \frac{\kappa}{2}\Vert \mathbf{P} - \mathbf{Q}\Vert_F^2.\nonumber
\label{eqn:nb11}
\end{eqnarray}
Note that this optimization problem has a closed-form solution, which can be expressed as:
\begin{eqnarray}
	\mathbf{Q} = (\frac{2\mathbf{Y}\mathbf{X}^T+\mathbf{T}}{\kappa})(\frac{\mathbf{X}\mathbf{X}^T}{\kappa}+\mathbf{I})^{-1},\nonumber
\label{eqn:nb12}
\end{eqnarray}
where $\mathbf{I}$ is a $d$-by-$d$ identity matrix.\\
(2) \textit{Fix $\mathbf{Q}$, $\mathbf{T}$, $\kappa$, and Minimize $\mathcal{L}$ w.r.t. $\mathbf{P}$}: In this step, we are required to solve the following optimization problem:
\begin{eqnarray}
	\min_{\mathbf{P}} Tr[\mathbf{T}^T(\mathbf{P} - \mathbf{Q})] + \frac{\kappa}{2}\Vert \mathbf{P} - \mathbf{Q}\Vert_F^2 +\lambda_3 \Vert \mathbf{P} \Vert_{2,1},\nonumber
\label{eqn:nb13}
\end{eqnarray}
which can be reformulated as follows:
\begin{eqnarray}
	\min_{\mathbf{P}} \frac{\lambda_3}{\kappa} \Vert \mathbf{P} \Vert_{2,1} + \frac12\Vert \mathbf{P} - (\mathbf{Q} - \frac{\mathbf{T}}{\kappa})\Vert_F^2.\nonumber
\label{eqn:nb13}
\end{eqnarray}

According to Lemma 4.1 shown in the work of~\cite{liu2012robust}, the optimal solution to the above optimization problem is
\begin{eqnarray}
	\mathbf{p}_i = \left\{
\begin{array}{lr}
\frac{\Vert \mathbf{p}_i - \frac{\mathbf{t}_i}{\kappa}\Vert - \frac{\lambda_3}{\kappa}}{\Vert \mathbf{q}_i - \frac{\mathbf{t}_i}{\kappa}\Vert}( \mathbf{q}_i - \frac{\mathbf{t}_i}{\kappa}), & \Vert \mathbf{q}_i - \frac{\mathbf{t}_i}{\kappa}\Vert> \frac{\lambda_3}{\kappa}, \\
\mathbf{0}, & otherwise.
\end{array}\right.\nonumber
\label{eqn:nb15}
\end{eqnarray}
where $\mathbf{p}_i$, $\mathbf{q}_i$, and $\mathbf{t}_i$ are the $i^{th}$ column of $\mathbf{P}$, $\mathbf{Q}$, and $\mathbf{T}$, respectively.\\
(3) \textit{Update $\mathbf{T}$ and $\kappa$}: 
\begin{eqnarray}
	\mathbf{T} = \mathbf{T} + \kappa(\mathbf{P} - \mathbf{Q}), \kappa = \min(\rho\kappa,\kappa_{max}),\nonumber
\end{eqnarray}
where $\rho>1$ and $\kappa_{max}$ is the preset maximal value for $\kappa$.\\
(4) \textit{Check Convergence}:
\begin{eqnarray}
	\Vert \mathbf{P} - \mathbf{Q} \Vert_F < \epsilon,\nonumber
\end{eqnarray}
where $\epsilon$ is the machine epsilon value.
\end{algorithm}

\subsection{Optimization of AKTLR}

The optimization problem for training AKTLR, as shown in Eq.(\ref{eqn:nb4}), can be effectively addressed using the alternated direction method (ADM)~\cite{zheng2014multi}. Specifically, the optimal parameters in AKTLR, represented by $\hat{\mathbf{P}}_i$ and $\hat\alpha$, can be obtained through the following iterative steps:

(1) \textit{Fix} $\mathbf{P}_i$ \textit{and Update} $\alpha$: In this step, the optimization problem becomes one with respect to $\alpha$, which can be formulated as follows:
\begin{eqnarray}
	\min_{\alpha}\Vert \mathbf{Y}^s - \sum_{i=1}^{G}\alpha_i\mathbf{P}_i\mathbf{X}_i^{s}\Vert_F^2 + \lambda_1 \Vert\sum_{i=1}^{G}\alpha_i\mathbf{P}_i\Delta\bar{\mathbf{x}}_i^{st}\Vert^2~~~~\nonumber\\+ \lambda_2 \Vert \alpha \Vert_1, \nonumber\\ s.t.~\alpha \succeq 0.~~~~~~~~~~~~~~~~~~~~~~~~~~~~~~~~~~~~~~~~~~~~~~~~~~
\label{eqn:nb5}
\end{eqnarray}
Let $\mathbf{Y} = [\mathbf{Y}^s, \mathbf{0}]$ and $\tilde{\mathbf{X}}_i = [\mathbf{X}_i^s, \sqrt{\lambda_1}\Delta\bar{\mathbf{x}}_i^{st}]$, where $\mathbf{0} \in \mathbb{R}^{C\times1}$ is a vector of all zero values. Then, the optimization problem in Eq.(\ref{eqn:nb5}) can be rewritten as:
\begin{eqnarray}
	\min_{\alpha}\Vert \mathbf{Y} - \sum_{i=1}^{G}\alpha_i\mathbf{P}_i\tilde{\mathbf{X}}_i\Vert_F^2 + \lambda_2 \Vert \alpha \Vert_1.
\label{eqn:nb6}
\end{eqnarray}

Subsequently, let $\mathbf{z}_i = Flatten(\mathbf{P}_i\tilde{\mathbf{X}}_i)~(i=\{1,\cdots,G\})$ and $\mathbf{y} = Flatten(\mathbf{Y})$, where $Flatten(\cdot)$ is an operation that reshapes a matrix into a vector column by column. We are thus able to further restate the objective function in Eq.(\ref{eqn:nb6}) as the following formulation:
\begin{eqnarray}
	\min_{\alpha}\Vert \tilde{\mathbf{y}} - \mathbf{Z}\alpha\Vert^2 + \lambda_2 \Vert \alpha \Vert_1,~s.t.~\alpha \succeq 0,
\label{eqn:nb7}
\end{eqnarray}
where $\mathbf{Z} = [\mathbf{z}_1,\cdots,\mathbf{z}_G]$. It is apparent that Eq.(\ref{eqn:nb7}) represents a standard non-negative LASSO problem, and we utilize the SLEP package~\cite{liu2009slep} to solve it.

(2) \textit{Fix} $\alpha$ \textit{and Update} $\mathbf{P}_i$: The optimization problem in this step can be formulated as follows:
\begin{eqnarray}
	\min_{\mathbf{P}}\Vert \mathbf{Y} - \mathbf{P}\mathbf{X}\Vert_F^2 + \lambda_3 \Vert \mathbf{P} \Vert_{2,1},
\label{eqn:nb8}
\end{eqnarray}
where $\mathbf{X} = [\mathbf{X}_1^T,\cdots,\mathbf{X}_G^T]^T~(\mathbf{X}_i = \alpha_i\tilde{\mathbf{X}}_i)$. We use the inexact augmented Lagrangian multiplier (IALM)~\cite{lin2010augmented} to learn the optimal $\mathbf{P}_i$. 

To be specific, an additional variable, $\mathbf{Q}$ satisifying $\mathbf{P} = \mathbf{Q}$, is introduced to first convert the original unconstrained optimization problem in Eq.(\ref{eqn:nb8}) to a constrained one, which can be expressed as follows:
\begin{eqnarray}
	\min_{\mathbf{P},\mathbf{Q}}\Vert \mathbf{Y} - \mathbf{Q}\mathbf{X}\Vert_F^2 + \lambda_3 \Vert \mathbf{P} \Vert_{2,1},~s.t.~\mathbf{P} = \mathbf{Q}.
\label{eqn:nb9}
\end{eqnarray}

Subsequently, we are able to obtain the Lagrangian function for Eq.(\ref{eqn:nb9}), which is formulated as follows:
\begin{eqnarray}
	\mathcal{L}(\mathbf{P},\mathbf{Q},\mathbf{T},\kappa) = \Vert \mathbf{Y} - \mathbf{Q}\mathbf{X}\Vert_F^2 + Tr[\mathbf{T}^T(\mathbf{P} - \mathbf{Q})] ~~~~~\nonumber\\+ \frac{\kappa}{2}\Vert \mathbf{P} - \mathbf{Q}\Vert_F^2 +\lambda_3 \Vert \mathbf{P} \Vert_{2,1},
\label{eqn:nb10}
\end{eqnarray}
where $\mathbf{T}$ is the Lagrangian multiplier matrix, $Tr(\cdot)$ represents the trace of a square matrix, and $\kappa$ is a relaxation factor.

Finally, the optimal solution of $\mathbf{P}$ can be obtained by iteratively minimizing the Lagrangian function in Eq.(\ref{eqn:nb10}) with respect to one of variables while fixing the others. The detailed updating procedures are summarized in Algorithm~\ref{alm:nb1}.

(3) \textit{Check Convergence}: the value of objective function is less than the machine epsilon value $\epsilon$ or that the iteration reaches the preset maximal number.

\subsection{Prediction of Emotion Labels for Target Speech Signals}

Once we have obtained the optimal solution, $\hat{\mathbf{P}}_i$ and $\hat\alpha$, for AKTLR, we can easily predict the emotion labels of the target speech signals. Let $\mathbf{x}^t = [{\mathbf{x}^t_1}^T,\cdots,{\mathbf{x}^t_G}^T]^T$ be the feature vector of a target speech sample. We first predict its emotion label vector $\hat{\mathbf{y}}^t$ by solving the following optimization problem:
\begin{eqnarray}
	\min_{\mathbf{y}^t}\Vert \mathbf{y}^t - \sum_{i=1}^{G}\hat\alpha_i\hat{\mathbf{P}}_i\mathbf{x}_i^{t}\Vert_F^2,~s.t.,~\mathbf{y}^t \succeq 0,~\mathbf{1}^T\mathbf{y}^t = 1.
\label{eqn:nb11}
\end{eqnarray}

This is a standard quadratic programming problem and can be effectively solved using the interior point method. Then, based on $\hat{\mathbf{y}}^t$, the emotion label of its corresponding target speech signal can be determined as the $j^{th}$ emotion, which satisfies the following criterion:
\begin{eqnarray}
	j = \arg\max_{j} \{\hat{\mathbf{y}}^t(j)~\vert~j=1,\cdots,C\},
\label{eqn:nb12}
\end{eqnarray}
where $\hat{\mathbf{y}}^t(j)$ represents the $j^{th}$ entry in the predcted emotion label vector $\hat{\mathbf{y}}^t$.

\section{Experiments}
\label{sec:experiments}
\subsection{Experiment Setup}

In this section, we evaluate the performance of the proposed AKTLR method through extensive cross-corpus SER experiments. We provide details of our experiment setup, including: 1) \textit{Speech Emotion Corpora}, 2) \textit{Experimental Protocol}, 3) \textit{Performance Metric}, and 4) \textit{Comparison Methods and Implementation Details}.

\subsubsection{Speech Emotion Corpora}

We utilize three publicly available speech emotion corpora in our experiments. Here is a brief overview of these corpora:

\textit{EmoDB}~\cite{burkhardt2005database}: This German speech emotion corpus consists of 535 speech samples. Each sample corresponds to a sentence uttered in German under one of seven emotional states (\textit{Anger}, \textit{Boredom}, \textit{Disgust}, \textit{Fear}, \textit{Happiness}, \textit{Neutral}, and \textit{Sadness}) by one of 10 professional German actresses/actors (five actresses and five actors).

\textit{eNTERFACE}~\cite{martin2006enterface}: Unlike EmoDB, eNTERFACE is a bimodal emotion database containing 1,257 video clips with both speech and facial expressions. Each video clip is labeled with one of six basic emotions (\textit{Anger}, \textit{Disgust}, \textit{Fear}, \textit{Happiness}, \textit{Sadness}, and \textit{Surprise}). For the design of our cross-corpus SER tasks, only the speech data is used.

\textit{CASIA}~\cite{zhang2008design}: This is a large-scale Chinese speech emotion corpus comprising 9,600 speech samples. In our experiments, we utilize its freely released version, which includes 1,200 speech samples from four speakers (two females and two males), with each speech sample conveying one of six different emotions (\textit{Anger}, \textit{Fear}, \textit{Happiness}, \textit{Neutral}, \textit{Sadness}, and \textit{Surprise}).

\subsubsection{Experimental Protocol}

We used the aforementioned three speech emotion corpora to create six cross-corpus SER tasks: $B\rightarrow E$, $E\rightarrow B$, $B\rightarrow C$, $C\rightarrow B$, $E\rightarrow C$, and $C\rightarrow E$. Here, $B$, $E$, and $C$ represent EmoDB, eNTERFACE, and CASIA, respectively. The corpora listed on either side of the arrow indicate the source and target speech emotion corpora in their respective cross-corpus SER tasks. It is important to note that due to inconsistencies in emotion labels across the three speech emotion corpora, only speech samples with matching emotion labels are chosen for their corresponding tasks. For a more comprehensive understanding of these cross-corpus SER tasks, detailed data composition for all the speech emotion corpora is presented in Table~\ref{tab:nb1}.

\begin{table*}[t]
\centering
\caption{Detailed Sample Composition for All Three Speech Emotion Corpora Used in the Experiments.}
\renewcommand{\arraystretch}{1.3}
\begin{tabular}{|cl|c|c|c|c|c|c|}
\hline
\multicolumn{2}{|c|}{\multirow{2}{*}{\textbf{Cross-Corpus SER Task}}} & \multicolumn{2}{c|}{\textbf{B$\rightarrow$E} / \textbf{E$\rightarrow$B}} & \multicolumn{2}{c|}{\textbf{B$\rightarrow$C} / \textbf{C$\rightarrow$B}} & \multicolumn{2}{c|}{\textbf{E$\rightarrow$C} / \textbf{C$\rightarrow$E}} \\ \cline{3-8} 
\multicolumn{2}{|c|}{} & ~~~\textbf{EmoDB}~~~ & \textbf{eNTERFACE} & ~~~\textbf{EmoDB}~~~ & ~~~~\textbf{CASIA}~~~~ & \textbf{eNTERFACE} & ~~~~\textbf{CASIA}~~~~ \\ \hline\hline
\multicolumn{1}{|c|}{\multirow{7}{*}{Sample Number}} & \textit{Anger} & 127 & 211 & 127 & 200 & 211 & 200 \\ \cline{2-8}
\multicolumn{1}{|c|}{}    & \textit{Fear} & 69 & 211 & 69 & 200 & 211 & 200 \\ \cline{2-8}
\multicolumn{1}{|c|}{}    & \textit{Disgust} & 46 & 211 & - & - & - & - \\ \cline{2-8}
\multicolumn{1}{|c|}{}    & \textit{Happiness} & 71 & 208 & 71 & 200 & 208 & 200 \\ \cline{2-8}
\multicolumn{1}{|c|}{}    & \textit{Neutral} & - & - & 79 & 200 & - & - \\ \cline{2-8}
\multicolumn{1}{|c|}{}    & \textit{Sadness} & 62 & 211 & 62 & 200 & 211 & 200 \\ \cline{2-8}
\multicolumn{1}{|c|}{}    & \textit{Surprise} & - & - & - & - & 211 & 200 \\ \hline\hline
\multicolumn{2}{|c|}{Total Number} & 375 & 1,052 & 408 & 1,000 & 1,052 & 1,000 \\
\hline
\end{tabular}
\label{tab:nb1}
\end{table*}

\subsubsection{Performance Metric}

We have chosen the unweighted average recall (UAR)~\cite{schuller2010cross} as the performance metric for our experiments. UAR is computed by averaging the accuracy across the total number of emotion classes. It is calculated using the formula \textit{UAR} $ = \frac{1}{C}\sum_{i=1}^{C} \frac{N_{i}^{p}}{N_{i}^{g}}\times100$. Here, $C$ is the number of total emotion classes involved in the cross-corpus SER task, and $N_{i}^{p}$ and $N_{i}^{g}$ represent the number of samples predicted as the $i^{th}$ emotion and the actual number of $i^{th}$ emotion samples, respectively.

\subsubsection{Comparison Methods and Implementation Details}

To highlight the effectiveness and superior performance of our AKTLR method in addressing the challenge of cross-corpus SER, we compare it with five recent state-of-the-art (SOTA) \textit{Transfer Subspace Learning} methods and six SOTA \textit{Deep Transfer Learning} methods. The methods included in the comparison and their implementation details are as follows:

\textit{Transfer Subspace Learning Methods} include transfer component analysis (TCA)~\cite{pan2010domain}, geodesic flow kernel (GFK)~\cite{gong2012geodesic}, subspace alignment (SA)~\cite{fernando2013unsupervised}, domain-adaptive subspace learning (DoSL)~\cite{liu2018unsupervised}, and joint distribution adaptive regression (JDAR)~\cite{zhang2021cross}. In these methods, two widely-used acoustic parameter feature sets, namely INTERSPEECH 2009 Emotion Challenge (IS09)~\cite{schuller2009interspeech} and the extended Geneva minimalistic acoustic parameter set (eGeMAPS)~\cite{eyben2015geneva}, are utilized to describe speech signals. Both feature sets consist of low-level descriptors (LLDs) such as F0 and MFCC through typical statistical functions. The openSMILE toolkit~\cite{eyben2010opensmile} is used to extract these feature sets from the speech signals. For the experiments, linear support vector machine (SVM)~\cite{chang2011libsvm} is used as the classifier for all subspace learning methods without classification ability, including TCA, GFK, and SA. Additionally, the results of directly using SVM to conduct all cross-corpus SER experiments are included as the baseline.

Since emotion label information is unavailable in the tasks of cross-corpus SER, we follow the tradition of transfer learning evaluation. Therefore, we report the best results of the five transfer subspace learning methods by searching their hyper-parameters from a given interval. Specifically, TCA, GFK, and SA aim to learn a $d$-dimensional common subspace for both source and target speech samples, where $d$ is set within a predetermined parameter interval, $[1:d_{max}]$, and $d_{max}$ represents the number of elements in the acoustic parameter set used in the experiments. DoSL and JDAR require setting two trade-off parameters, $\lambda$ and $\mu$, which control the balance between the sparsity and feature distribution elimination terms and the original regression loss function. In the experiments, $\lambda$ and $\mu$ are determined by searching within the range of $[1:100]$.

\textit{Deep Transfer Learning Methods} including deep adaptation network (DAN)~\cite{long2015learning}, joint adaptation network (JAN)~\cite{long2017deep}, deep subdomain adaptation network (DSAN)~\cite{zhu2020deep}, domain-adversarial neural network (DANN)~\cite{ajakan2014domain}, conditional domain adversarial network (CDAN)~\cite{long2018conditional}, and DIDAN~\cite{zhao2023deep}, are utilized in the comparison experiments. The speech signals are first tranformed into the Mel-spectrograms and then resized to $224 \times 224$ pixels, serving as the input for deep neural networks. In this comparison, VGG-11~\cite{simonyan2014very} is chosen as the CNN backbone of all the deep transfer learning methods, and its experimental results are included as the baseline. The optimizer, learning rate, weight decay, and batch size are set as SGD, $1e^{-2}$, $5e^{-4}$, and $32$, respectively, for the VGG-11 and comparison deep transfer learning methods. The trade-off parameter settings for all deep transfer learning methods are as follows:

DAN, JAN, DSAN, DANN, and CDAN have a trade-off parameter $\lambda$ in their loss functions, which balances the original loss function and the feature distribution alleviation term. In the experiments, $\lambda$ is searched within the parameter interval $[0.0001:0.0001:0.001,0.002:0.001:0.01,0.02:0.01:0.1,0.2:0.1:1,2,5,10,100]$. Besides $\lambda$, \textit{DIDAN} has an additional trade-off parameter, $\alpha$, which controls the sparsity of its learned reconstruction coefficient matrix. For DIDAN, $\lambda$ and $\alpha$ are also searched within the same intervals as the other five deep transfer learning methods: $[0.0001:0.0001:0.001,0.002:0.001:0.01,0.02:0.01:0.1,0.2:0.1:1,2,5,10,100]$.

Our AKTLR has three trade-off parameters: $\lambda_1$ and $\lambda_3$. In our experments, we conduct a search for $\lambda_1$ and $\lambda_3$ in the parameter interval of $[1:100]$, while $\lambda_2$ is searched winthin the range of $[0.1:0.1:1]$. Additionally, we divide both IS09 and eGeMAPS feature sets into 10 LLD groups based on the acoustic parameter type. For further details, please refer to Table~\ref{tab:nb2}. 

\begin{table}[t!]
	\centering
	\renewcommand{\arraystretch}{1.3}
	\caption{Configuration of LLD Groups for AKTLR Using IS09 and eGeMAPS Feature Sets to Describe Speech Signals (element numbers are indicated in parentheses).}
	\begin{tabular}{|c|c|}
		\hline
		\textbf{Feature Set} &  \textbf{LLD Groups}\\
		\hline\hline
		\multirow{3}*{IS09} & ZCR (12), $\Delta$ZCR (12), F0 (12), $\Delta$F0 (12), \\
& RMS Energy (12), $\Delta$RMS Energy (12), HNR (12),   \\
                                    & $\Delta$HNR (12), MFCC (144), $\Delta$MFCC (144)\\\hline
		\multirow{4}*{eGeMAPS}   & F0 (18), Loudness (16), Spectral Flux (5), \\ 
									&Formant (18), Hammarberg Index (3), MFCC (16), \\
									& Spectral Slope (6), Alpha Ratio (3), HNR (2), \\
									& Equivalent Sound Level (1)\\ \hline		  
	\end{tabular}
	\label{tab:nb2}
\end{table}

\begin{table*}[t!]
	\centering
	\renewcommand{\arraystretch}{1.3}
	\caption{Comparison of the Proposed AKTLR Method and Recent State-of-the-Art Transfer Learning Methods for Cross-Corpus SER Tasks. The Best Result in Each Task is Highlighted in Bold.}
	\begin{tabular}{|c|c|c|c|c|c|c|c|c|}
		\hline
		\multicolumn{2}{|c|}{\textbf{Method}} & $\textbf{B}\rightarrow \textbf{E}$ & $\textbf{E}\rightarrow \textbf{B}$ & $\textbf{B}\rightarrow \textbf{C}$ & $\textbf{C}\rightarrow \textbf{B}$ & $\textbf{E}\rightarrow \textbf{C}$ & $\textbf{C}\rightarrow \textbf{E}$ & \textbf{Average}\\
		\hline \hline
		\multirow{6}{*}{\makecell[c]{Subspace Learning \\ (IS09 Feature Set)}}
&SVM&~~28.93~~&~~23.58~~&~~29.60~~&~~35.01~~&~~26.10~~&~~25.14~~&~~28.06~~\\
		& TCA & 30.73 & 45.16 & 33.40 & 45.82 & 31.80 & 34.12 & 36.84\\
		& GFK & 32.40 & 45.42 & 35.60 & 51.19 & 32.90 & 29.54 & 37.84\\
		& SA & 33.50 & 45.78 & 36.90 & 48.48 & 32.80 & 32.71 & 38.36\\
		& DoSL & 36.29 & 39.84 & 34.60 & 46.14 & 30.90 & 31.69 & 36.58\\
		& JDAR & 37.10 & 40.78 & 33.10 & 47.34  & 32.40 & 31.50 & 37.04 \\ \hline
		\multirow{6}{*}{\makecell[c]{Subspace Learning \\ (eGeMAPS Feautre Set)}}
 		& SVM & 25.65 & 32.58 & 33.50 & 51.84 & 36.40 & 34.79 & 35.96 \\
		& TCA & 31.09 & 37.43 & 42.90 & 53.43 & \textbf{41.10} & \textbf{35.90} & 40.31\\
		& GFK & 30.08 & 35.79 & 40.00 & 50.79 & 39.20 & 34.48 & 38.39\\
		& SA & 32.18 & 39.37 & 38.80 & 53.20 & 37.00 & 35.43 & 39.33\\
		& DoSL & 30.81 & 40.71 & 39.30 & 52.21 & 39.10 & 34.27 & 39.40\\
		& JDAR & 31.41 & 45.19 & 42.30 & 56.14  & 38.40 &33.62 & 41.18\\  \hline \hline
		\multirow{7}{*}{Deep Learning} & VGG-11 &  27.08 & 34.83  & 34.80  & 51.31 &26.90&26.02&33.49 \\
		& DAN & 33.58 & 43.50&36.30 & 56.72 & 29.30 & 32.17 & 38.60 \\
		& JAN & 35.23 & 47.29 & 37.00 & 57.51 & 31.00 & 32.21 & 40.04 \\
		& DSAN & 31.82 & \textbf{47.58} & 35.58 & 56.50 & 29.00 & 31.25 & 38.66 \\
		& DANN & 32.56 & 46.06 & 36.40 & 57.67 & 30.50 & 33.77 & 39.49 \\
        & CDAN & 31.62 & 46.12 & 35.40 & 57.60 &30.30 & 33.49 & 39.09 \\
        & DIDAN & 33.05 & 47.11 & 38.90 & 56.22 & 31.10 &34.06&40.07 \\
  \hline\hline
		\multirow{2}{*}{Subspace Learning} & \textbf{AKTLR (IS09)} & \textbf{37.51}  &  47.12 &  37.00  & 47.61  &  30.60 &  33.11  & 38.83\\
		&  \makecell[c]{\textbf{AKTLR (eGeMAPS)}} & 32.51 & 43.60& \textbf{45.00} & \textbf{59.93}  & 37.60  & 34.09 & \textbf{42.12}\\ \hline
	\end{tabular}
	\label{tab:nb3}
\end{table*}

\subsection{Comparison with State-of-the-Art Cross-Corpus SER Methods}
\label{sec:comparison}

The experimental results for all transfer learning methods are presented in Table~\ref{tab:nb3}. Several noteworthy observations can be made from this table:

(1) It is evident from Table~\ref{tab:nb3} that both transfer subspace learning and deep transfer learning methods exhibit promising performance improvements compared to their respective baseline methods (SVM or VGG-11) in all six cross-corpus SER tasks. Particularly interesting is the consistent enhancement observed in transfer subspace methods, regardless of the choice of acoustic parameter feature sets (IS09 or eGeMAPS) used to describe speech signals. In summary, our experimental results strongly indicate the potential of transfer learning as a promising approach to effectively address the challenge of cross-corpus SER.

(2) The performance comparison of transfer subspace learning methods using the IS09 (16 LLDs yielding 384 features) and eGeMAPS feature sets (five meticulously chosen LLDs yielding 88 features) reveals that the eGeMAPS feature set significantly improves cross-corpus SER performance compared to IS09. This finding underscores the importance of selecting minimalistic high-quality acoustic parameters capable of exhibiting superior generalization ability to corpus invariance when employing transfer subspace learning methods to address cross-corpus SER tasks. Our results provide additional experimental evidence to support this established knowledge in SER~\cite{williams1972emotions,parlak2014cross,eyben2015geneva}, which motivates the design of our AKTLR method.

(3) As shown in the table, our AKTLR, utilizing the eGeMAPS feature set, achieves the highest UAR among all transfer learning methods, averaging a UAR of $42.12\%$ across the six cross-corpus SER tasks. Furthermore, our AKTLR outperforms all other methods in two out of the six tasks, namely $B\rightarrow C$ and $C\rightarrow B$. While AKTLR may not achieve the best performance in the remaining four tasks, it still demonstrates a very competitive performance compared to all other transfer learning methods. In summary, these observations highlight the superior performance of our AKTLR method in addressing the challenge of cross-corpus SER, surpassing both recent SOTA transfer subspace learning and deep transfer learning methods. This also demonstrates the feasibility and superiority of incorporating acoustic knowledge to develop a domain-specific cross-corpus SER approach for dealing with cross-corpus SER tasks.

\begin{table}[!t]
	\centering
	\renewcommand{\arraystretch}{1.3}
	\caption{Detailed Configuration of Additional LLD Group Settings for eGeMAPS Feature Set. The Number of Elements is Given in Parentheses.}
	\begin{tabular}{|c|c|}
		\hline
		\textbf{\#LLD Groups} &  \textbf{Details of LLD Groups}\\\hline\hline
		\multirow{2}*{4 Groups} & Frequency (30), Energy (20), \\
								  &Spectral (37), Equivalent Sound Level (1) \\ \hline
		\multirow{5}*{13 Groups} & F0 (10), Jitter (2), Formant (18), Spectral Slope (6), \\
									&  MFCC (16),  Alpha Ratio (3), Shimmer (2), \\
									& Hammarberg (3), HNR (2), Harmonic Difference (4), \\
									&  Spectral Flux (5), MFCC (16), Londness (16), \\
									& Equivalent Sound Level (1) \\ \hline
	\end{tabular}
	\label{tab:nb4}
\end{table}

\subsection{A Deeper Look at the Proposed AKTLR Method}

This section aims to provide a comprehensive understanding of the proposed AKTLR method. We will address three key questions to delve into AKTLR: 1) Does AKTLR truly benefit from the incorporation of the selected acoustic knowledge? 2) What can AKTLR learn guided by the selected acoustic knowledge? 3) How does the performance of AKTLR vary with changes in the trade-off parameter?. To answer these questions, we will conduct additional cross-corpus SER experiments using AKTLR, with the aim of offering comprehensive insights into its effectiveness and advantages.

\begin{table}[t!]
	\centering
	\renewcommand{\arraystretch}{1.3}
	\caption{The experimental results of state-of-the-art methods using eGeMAPS feature set on six cross-corpus SER tasks. (\%)}
	\begin{tabular}{|l|ccc|}
		\hline
		\textbf{Method} & $\textbf{B}\rightarrow \textbf{E}$  & $\textbf{B}\rightarrow \textbf{C}$ &  $\textbf{E}\rightarrow \textbf{C}$  \\
		\hline\hline
		AKTLR w/o $\Vert\alpha\Vert_1$ (No Group) & ~~30.08~~ & ~~39.90~~ & ~~39.10~~  \\\hline\hline
		AKTLR (4 Groups) &  \textbf{33.29} & 43.40 & \textbf{39.70} \\
		AKTLR (10 Groups) & 32.51 & \textbf{45.00} & 37.60\\ 
		AKTLR (13 Groups) & 32.51 & 42.50 & 37.60\\ \hline
	\end{tabular}
	\label{tab:nb5}
\end{table}

\begin{figure*}[t!]
\centering
\includegraphics[width=\textwidth]{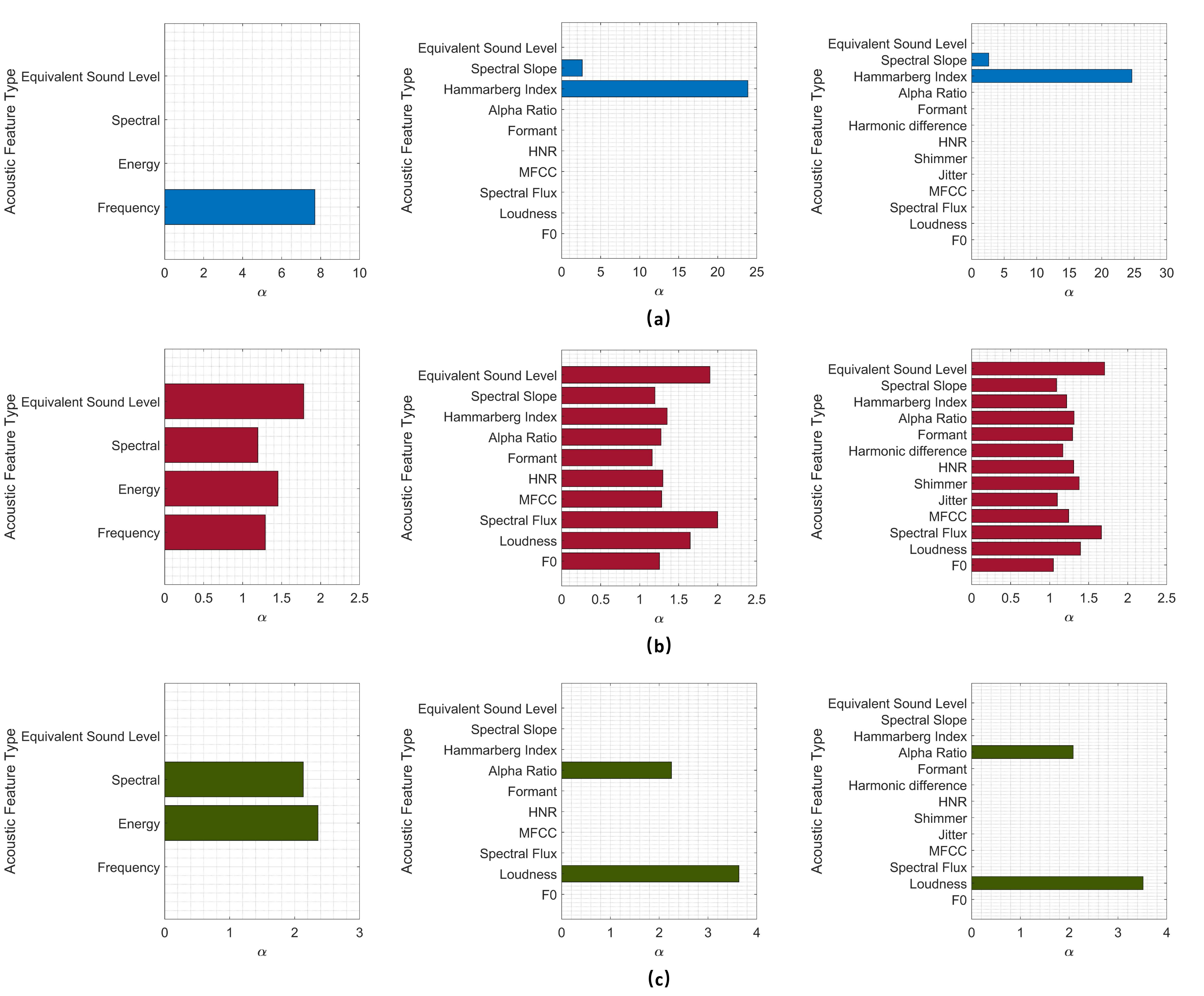}
\caption{The bar charts for the learned $\alpha$ by AKTLR, depicting the specific contributions of their corresponding acoustic parameter derived features for cross-corpus SER. (a), (b), and (c) correspond to Tasks $B\rightarrow E$, $B\rightarrow C$, and $E\rightarrow C$.}
\label{fig:nb2}
\end{figure*}

\subsubsection{Does AKTLR Truly Benefit From the Incorporation of the Selected Acoustic Knowledge?}
\label{sec:nbdeep1}

To address this question, we conduct additional experiments on three representative cross-corpus SER tasks: $B\rightarrow E$, $B\rightarrow C$, and $E\rightarrow C$. Specifically, we utilize the eGeMPAS feature set to describe speech signals, which is divided into two additional LLD groups: $G=4$ and $G=13$ for AKTLR, different from the previous experiments where $G = 10$. The detailed configuration of LLD group settings can be found in Table~\ref{tab:nb4}. In these experiments, we also remove the regularization term $\Vert \alpha \Vert_1$ from the objective function of AKTLR, resulting in a reduced version of AKTLR that alighs with the objective function of DoSL~\cite{liu2018unsupervised}. Thus, this reduced version can be viewed as AKTLR without specially considering the different contributions of LLDs, denoted as AKTLR w/o $\Vert \alpha \Vert_1$ (No Group). The experimental results, presented in Table~\ref{tab:nb5}, reveal several interesting observations that provide an experimental answer to this question.

\begin{figure*}[!t]
\centering
\includegraphics[width=\textwidth]{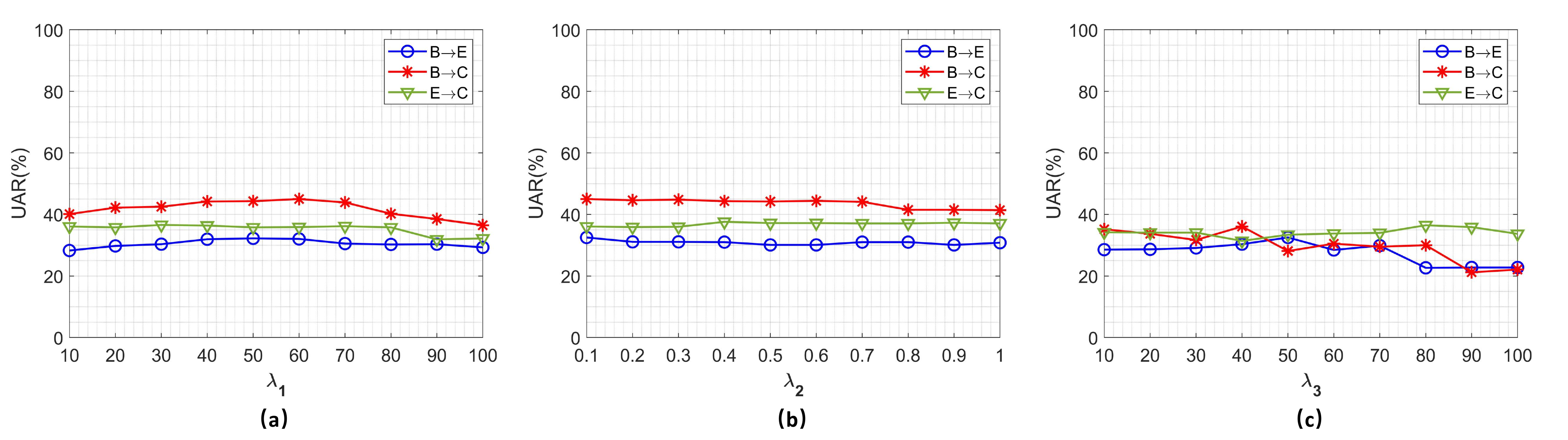}
\caption{The experimental results of trade-off parameter sensitivity analysis for our proposed AKTLR in addressing the tasks of cross-corpus SER, where (a), (b), and (c) correspond to the results of changing $\lambda_1$, $\lambda_2$, and $\lambda_3$ while fixing others.}
\label{fig:nb3}
\end{figure*}

Firstly, it is evident that our AKTLR models, which adopt different LLD group settings, achieve better performance in terms of UAR compared to AKTLR without setting LLD groups. This observation demonstrates the feasibility and superiority of the concept behind our proposed AKTLR, i.e., "selecting these acoustic parameters may enable the transfer subspace learning models to achieve more promising recognition performance in cross-corpus SER tasks compared to directly using larger feature sets comprising comprehensive acoustic parameters". Guided by this acoustic knowledge, AKTLR divides the acoustic parameter feature set into different LLD groups and measures their contribution scores, ensuring the learning of both emotion-discriminative and corpus-invariant features.

Secondly, it is worth noting that the AKTLR models with 10 and 13 groups perform worse than AKTLR without without setting LLD groups in the task of $E\rightarrow C$. We believe that this is mainly due to the use of an excessive LLD groups in these cases. By comparing the different groups used for various cross-corpus SER tasks, it becomes apparent that the overall performance of AKTLR decreases with an increase in the number of groups. This supports our previous supposition. In other words, determining a suitable LLD group setting remains an open question for our AKTLR method in tackling the challenge of cross-corpus SER.

\subsubsection{What Can AKTLR Learn When Guided by the Selected Acoustic Knowledge?}

Our proposed AKTLR benefits from the incorporation of established acoustic knowledge into its design. By dividing the acoustic parameter feature set into different LLD groups and measuring their contribution scores, AKTLR model is more capable of seeking a minimalistic high-quality features that are emotion-discriminative features and corpus-invariant. This approach inspires us to explore what AKTLR can learn when guided by the utilization of acoustic knowledge. To this end, we present a set of bar charts in Fig.~\ref{fig:nb2}, illustrating the $\alpha_i$ values learned by AKTLR when utilizing the eGeMAPS feature set with different LLD groups to address three representative cross-corpus SER experiments in Table~\ref{tab:nb5}.

The findings from Fig.~\ref{fig:nb2} are quite intriguing. Firstly, it is evident that different LLD groups exhibit varying contributions when addressing cross-corpus SER tasks. Specifically, in five out of the nine cross-corpus SER experiments, certain acoustic parameters (corresponding to 0-valued $\alpha_i$) show negligible contribution in distinguishing emotions across speech corpora. These observations provide experimental evidence that supports selected acoustic knowledge guiding the design of the proposed AKTLR~\cite{williams1972emotions,parlak2014cross,eyben2015geneva}. This implies that selecting minimalistic high-quality acoustic parameters is necessary and sufficient for dealing with the cross-corpus SER tasks.

Secondly, upon further examination of the contributive LLD groups, it becomes apparent that the contributions of several acoustic parameters vary across different cross-corpus SER tasks, exhibiting high scores in some tasks while low scores in others. This suggests that there are no consistently highly-contributive acoustic parameters for all the cross-corpus SER tasks. However, it is interesting to note the presence of several "stable" (varied but consistently contributive) emotion-discriminative and corpus-invariant acoustic parameters, such as MFCC, which consistently exhibit a satisfactory learned score. This insight inspires us to consider the possibility of testing and selecting acoustic parameters to develop a general minimalistic acoustic parameter feature set consisting of high-quality elements that are consistently emotion-discriminative and corpus-invariant. Such a set could potentially enhance the performance of transfer learning methods in addressing the challenge of cross-corpus SER.

\subsubsection{How Trade-off Parameters Affect the Performance of AKTLR?}

In Eq.(\ref{eqn:nb4}), our AKTLR requires to set three trade-off parameters: $\lambda_1$, $\lambda_2$, and $\lambda_3$. This raises the question of how the choice of these trade-off parameters affect the performance of AKTLR in addressing the challenge of cross-corpus SER. To investigate this point, we continue to conduct experiments using the eGeMAPS feature set on three cross-corpus SER tasks chosen above: $B\rightarrow E$, $B\rightarrow C$, and $E\rightarrow C$ . We change the value of one trade-off parameter while keeping the others fixed, and monitor the experimental results of AKTLR. The intervals for the trade-off parameter values are set as $[10:10:100]$ for both $\lambda_1$ and $\lambda_3$, and $[0.1:0.1:1]$ for $\lambda_2$. The fixed values for $\lambda_1$, $\lambda_2$, and $\lambda_3$ are those used in the experiments described in Section~\ref{sec:comparison}.

The results are illustrated in Figure \ref{fig:nb3}. From this figure, it is evident that the performance of our AKTLR varies slightly with respect to the choice of $\lambda_1$ and $\lambda_2$ across all three cross-corpus SER tasks. However, in the case of $\lambda_3$, although the performance of AKTLR appears to be sensitive to changes in its value, AKTLR consistently performs within an acceptable range around the fixed value used in the experiments. In summary, we can conclude that the performance of our AKTLR is generally less sensitive to the choice of its associated trade-off parameters.

\section{Conclusion}
\label{sec:conclusion}

In this paper, we have addressed the challenge of cross-corpus SER from a new perspective by introducing a novel transfer subspace learning method called AKTLR. The primary contribution of AKTLR lies in its acoustic knowledge-guided dual sparsity constraint mechanism, which enables more effective learning of emotion-discriminative and corpus-invariant features at two different scales: acoustic parameter and feature. Compared with existing transfer subspace learning-based cross-corpus SER methods, AKTLR is the first domain-specific approach designed specifically under the guidance of established acoustic knowledge for cross-corpus SER. To evaluate the effectiveness of AKTLR, we conduct extensive cross-corpus SER experiments using three widely-used speech emotion corpora. The results demonstrate that AKTLR outperforms current SOTA transfer subspace learning and deep transfer learning-based cross-corpus SER methods. This confirms the efficacy and feasibility of leveraging acoustic knowledge to develop domain-specific transfer learning methods for cross-corpus SER.


%

%
%
%
%
%

\ifCLASSOPTIONcaptionsoff
  \newpage
\fi



\bibliographystyle{IEEEtran}
\bibliography{TCSS2023}
\end{document}